# Disorder-enabled Synthetic Metasurfaces

Chi Li,[1,#] Changxu Liu,[2,#,*] Cade Peters[3], Haoyi Yu[1], Stefan A. Maier,[1,4] Andrew Forbes,[3]

Haoran Ren[1,*]

[1] School of Physics and Astronomy, Monash University, Melbourne, VIC, 3800, Australia

[2] Centre for Metamaterial Research & Innovation, Department of Engineering, University of Exeter, Exeter EX44QF, United Kingdom

[3] School of Physics, University of Witwatersrand, Johannesburg, South Africa

[4] Department of Physics, Imperial College London, London SW7 2AZ, United Kingdom

[#] These authors contributed equally

[*]To whom correspondence Emails:

C.C.Liu@exeter.ac.uk;

Haoran.ren@monash.edu.

**Abstract**

Optical metasurfaces have catalysed transformative advances across imaging, optoelectronics, quantum information processing, sensing, energy conversion, and optical computing. Yet, despite this rapid progress, most research remains focused on optimising single functionalities—constrained by the persistent challenge of integrating multiple functions within a single device. Here, we demonstrate that engineered structural disorder of meta-pixels—used to implement a photonic function—can significantly reduce the area required across the entire aperture without compromising optical performance. The unallocated space can then be repurposed to encode functionally distinct meta-pixels without increasing the design complexity, each independently addressable via various optical degrees of freedom. As a proof of concept, we present a synthetic achromatic metalens featuring 11 spectrally distinct lens profiles encoded through nonlocal meta-pixels engineered to support sharp resonances via quasi-bound states in the continuum. This large-scale metalens with 8.1 mm-aperture achieves diffraction-limited achromatic focusing across the 1200–1400 nm spectral window. We further incorporate polarisation-selective meta-pixels to implement momentum-space-distinct gratings, enabling single-shot, high-spatial-resolution polarimetric imaging of arbitrarily structured light fields, including radial and azimuthal vector beams and optical skyrmions. Altogether, this disorder-enabled synthetic metasurface platform establishes a versatile foundation for unifying diverse photonic functionalities within a single optical element, marking a substantial step toward compact, high-density, multifunctional optical devices.

Optical metasurfaces—planar arrays of subwavelength structures engineered to manipulate the phase, amplitude, and polarisation of light—have catalysed a paradigm shift in photonics, driving transformative advances in imaging [1, 2], optoelectronics [3], quantum information processing [4], sensing [5, 6], energy conversion [7], and optical computing [8-10]. Early developments centred on single-function metasurfaces, exemplified by optical beam steering [11, 12], diffraction-limited focusing lenses [13], and ultrathin holographic displays [14-18]. The increasing demand for compact, integrated photonic systems has propelled the field toward multifunctional metasurfaces. A prevalent approach harnesses meta-atom engineering, where geometrically tailored meta-atoms or supercells realise functionalities such as achromatic focusing [19-25], polarisation-sensitive imaging [26-28], and multiplexed displays [29, 30]. While promising, this strategy remains fundamentally constrained by the limited degrees of freedom available in meta-atom design, restricting both the diversity and complexity of achievable optical functionalities.

In parallel, the concept of floorplanning—a foundational principle in integrated circuit design—strategically optimises spatial allocation to balance global area efficiency, functional density ($D_f$, number of functions realised per unit area), and crosstalk suppression in electronic systems. This approach has enabled highly efficient electronic layouts, where engineered spatial disorder improves performance by optimising the placement of functional targets [31]. By contrast, efficient area utilisation in metasurface research remains largely unexplored. Existing approaches—such as segmentation (dividing the metasurface into discrete functional regions) [27, 28, 32, 33] and interleaving (alternating unit cells with distinct functions) [34]—have validated the feasibility of multifunctional metasurfaces. However, as functional density increases, both approaches encounter performance bottlenecks, suffering from partial beam coverage and/or lacking sufficient optical selectivity to maintain independent functionality. These limitations fundamentally restrict the number of realisable functions in a single metasurface device.

Disorder is typically perceived as detrimental in photonics, as random variations in shape, size, position, or orientation tend to disrupt optical symmetries and degrade device performance. Yet, when judiciously engineered, disorder has been harnessed to unlock unconventional functionalities—including random lasers [35], structural colour generation [36, 37] and broadband energy harvesting [38]. In the context of metasurfaces, engineered disorder [39] has likewise been exploited to suppress multiplexing crosstalk in 3D holography [40], enable vortex-[18] and polarisation-multiplexing holography [41], and facilitate the development of real-momentum-space topological states [42] and reconstructive spectrometers [43].

Here, we demonstrate that disorder can be strategically harnessed to enhance the functional density of metasurfaces, addressing the longstanding challenge of efficient area utilisation. Through a synthetic design strategy, we introduce randomised distributions of functionally distinct unit cells—termed meta-pixels (**Fig. 1a**)—which exhibit optically selective responses across multiple degrees of freedom, implemented via either local- or nonlocal-type phase pixels. Embedded within the metasurface aperture, these meta-pixels are encoded with distinct phase wavefronts that can be selectively accessed via wavelength, polarisation, and orbital angular momentum (OAM). By leveraging spatial disorder, the area required to realise each distinct function can be significantly reduced without compromising optical performance, laying the foundation for the seamless integration of multiple functionalities within a single metasurface aperture.

As a proof of concept, we demonstrate the integration of 11 spectrally distinct lens profiles—realised via nonlocal meta-pixels supporting quasi bound states in the continuum (qBIC)—all sharing identical focal lengths and collectively achieving achromatic light focusing (**Fig. 1b**).

Beyond lensing, this platform supports multifunctional displays of spectrally distinct OAM modes and holographic images (**Fig. 1c**). Furthermore, the incorporation of three sets of polarisation-selective meta-pixels—each implementing distinct diffraction gratings for orthogonal polarisation bases—enables previously unattainable single-shot, high-spatial-resolution polarimetric imaging of arbitrarily structured light fields, including radial and azimuthal vector beams and optical skyrmions (**Fig. 1d**). Overall, this disorder-enabled synthetic metasurface platform facilitates the integration of diverse photonic functions within a single optical element, substantially expanding the design landscape for high-density, multifunctional devices.

Traditionally, the entire area of a metasurface is dedicated to a single function. For instance, a metasurface lens (metalens) with a hyperbolic phase profile is fully employed to achieve light focusing. However, this design approach may involve a degree of redundancy, opening the possibility for re-evaluating the spatial resource allocation within the metalens to improve its functional density and versatility. To assess the minimum area required to achieve a given photonic function, we systematically investigate how partial utilisation of the metalens pixels affects its performance. Two types of configurations are considered: the first selects a continuous sector from the full aperture area of the metalens (ordered), while the second randomly selects active pixels across the metalens area (disordered). **Figure 2a** illustrates both configurations, where the red regions represent the utilised area and the black regions indicate unused portions. The parameter $p$ denotes the fraction of the total area that is employed. For simplicity, yet without loss of generality, we choose focusing at a single operational wavelength as the representative target function.

**Figures 2b** summarises the transverse point spread functions (PSFs) for both the ordered and disordered configurations as the area usage fraction $p$ gradually decreases. In the ordered case (left inset maps, upper row), we observe a prominent degradation in focusing performance: the focal spot becomes enlarged and distorted. In contrast, the disordered configuration (left inset maps, lower row) maintains a nearly unchanged focal distribution even when only 10% of the area is used. To quantitatively evaluate the lens performance, we employ the Strehl ratio (SR) as a figure of merit (more details in Supplementary Materials, S1A), with results presented as curves in blue (order) and red (disorder). Typically, an SR value of 1 indicates an aberration-free optical imaging system operating at the diffraction limit, while lower values reflect the presence of imperfections and aberrations within the system. For the ordered configuration, SR decreases continuously as the area usage is reduced. In contrast, the disordered configuration exhibits highly stable SR, with a noticeable drop only when $p < 0.1$. The right inset presents an extreme case where just 0.1% of the area is employed in the disordered case; despite some distortions, the light is still focused onto the designed focal region. Furthermore, at equal area usage fractions, disorder-enabled random sampling outperforms uniform grid sampling by suppressing additional diffraction orders associated with the Bragg diffraction (see Supplementary Materials, Section S2A and **Figure S1**). These findings underscore the remarkable robustness of disordered area selection. By leveraging spatial disorder, the required area for a given function can be significantly reduced, laying the foundation for integrating multiple functionalities into a single metasurface aperture. **Figure 2c** further demonstrates the order to disorder transition when half of the area is utilised ($p = 0.5$). To quantitatively characterise the degree of spatial order, we use Moran's spatial autocorrelation $I_m$, where $I_m = 1$ represents a fully ordered distribution and $I_m = 0$ corresponds to a fully randomised, minimum-clustering configuration [44] (more details in Supplementary Materials, S1B). As the configuration transitions from ordered to disordered, the SR dramatically improves from 0.6 to nearly 1.0—underscoring the unique capability of disorder to preserve optical performance under stringent area constraints. Further analysis of energy efficiency is provided in the Supplementary

Materials, S2B and **Fig. S2**.

To validate the concept, we numerically demonstrate that disorder-enabled functional density can be directly applied to the design of synthetic achromatic metalenses—combining a range of wavelength-specific lens phase functions sharing a common focal distance. In this framework, focusing at each wavelength is treated as an independent optical function. **Figure 2d** presents an example in which three such functions—corresponding to focusing at three distinct wavelengths—are synthesised into a single metasurface. The functional density is defined as $D_f = 1/p$. In this example, $p = 1/3$, indicating that each function occupies one-third of the metasurface area (highlighted in white), while the remaining regions (shown in black) are non-participating due to zero transmission. This underscores that sharp optical selectivity is crucial for isolating photonic functions during multifunctional operation, particularly since light of different wavelengths may impinge upon the same spatial locations.

For simplicity, we assume ideal spectral selectivity, wherein each lens function is exclusively activated by its corresponding wavelength. As illustrated in **Fig. 2e** (upper panel), each phase pixel is modelled with unitary transmission for its designated spectral band and zero transmission outside it. This assumption enables crosstalk-free synthesis of achromatic focusing across a target spectral range, governed by the $D_f$ and the bandwidth of each constituent function (**Fig. 2e**, lower panel). **Figure 2f** assesses the achromatic focusing performance by analysing the SR of the focal spot as a function of both $D_f$ and wavelength. As the number of functions increases, a broader spectral range is maintained with high SR values, resulting in enhanced achromaticity. Notably, even at high $D_f$ values (e.g., 41), where each function occupies only a small portion of the metasurface, the PSF remains well preserved (**Fig. S3**). In contrast to conventional achromatic designs [19-25, 45]—which are typically constrained in aperture size due to the limited design space for simultaneous phase and dispersion engineering—our approach places no such restriction, allowing for large-area devices with uncompromised focusing performance.

To experimentally realise the synthetic achromatic metalens, we employed T-shaped meta-pixels supporting qBIC nonlocal optical resonances. These meta-pixels simultaneously offer sharp wavelength selectivity and geometric phase control—exhibiting a $4\theta$ phase shift via in-plane rotation angle $\theta$ under circularly polarised illumination, owing to their two-fold rotational symmetry [46, 47]. This strong spectral selectivity satisfies the spatial isolation criterion for multifunctional synthesis, ensuring that each wavelength-specific lens function can be independently and resonantly addressed. We investigated the minimum number of unit cells per meta-pixel required to support pronounced qBIC (**Fig. S4**), which determines the spatial resolution of the synthetic metalens. T-shaped meta-pixels with varying unit cell numbers were fabricated in amorphous silicon (a-Si) and characterised using a transmission spectrometer system (see Methods, Supplementary Materials S3B, and **Fig. S5**). **Figure 3a** displays optical transmission images under white light illumination and cross-circular polarisation detection. From top to bottom, bright squares indicate efficient polarisation conversion from left- to right-handed circular polarisation, corresponding to a reduction in T-shaped unit cell numbers from $50 \times 50$ to $1 \times 1$ per meta-pixel. For simplicity, each square comprises two parallel T-shaped unit cells.

Spectral measurements (**Fig. 3b**) reveal clear resonant peaks emerging from the 6 x 6 unit cell configuration. The qBIC resonance functions as a spectral filter, exhibiting consistent spectral shape with increasing intensity for larger unit cell numbers. To realise wavelength-selective functionality, we applied a linear scaling factor to the transverse dimensions of the meta-pixels (**Fig. 3c**), enabling precise tuning of resonance wavelengths across 1200–1400 nm by adjusting the scaling factor from 0.7 to 1. For the synthetic achromatic metalens, we fixed the meta-pixel footprint at 8.1 µm × 8.1 µm, corresponding to 7 × 7 unit cells at 1400 nm, with larger unit cell numbers used for shorter wavelengths. Notably, resonance quality is well preserved during in-plane rotation of the T-shaped elements within each meta-pixel (**Fig. 3d**), supporting the full 0–2π geometric phase span, as indicated by the 0–π/2 in-plane rotations in the insets of the first column. Further analysis of resonance quality factors is provided in **Fig. S6**.

We designed and fabricated a synthetic achromatic metalens composed of 11 wavelength-specific shares, operating across a broad wavelength range from 1200 nm to 1400 nm. Optical images of the fabricated device are shown in **Fig. 4a**. The metalens comprises 1000 × 1000 meta-pixels, each with a fixed footprint of 8.1 µm x 8.1 µm, yielding an overall aperture diameter of approximately 8 mm (left panel). Under resonant excitation, meta-pixels corresponding to different shares become distinctly visible, as shown in the high- magnification dark-field optical image (right panel). A magnified view of these meta-pixels is shown in the false-colour scanning electron microscope (SEM) image in **Fig. 4b**. Sub-micrometre gaps between adjacent meta-pixels arise from the varying number of unit cells used to define each wavelength-specific meta-pixel.

To benchmark focusing performance, we first fabricated and characterised a reference metalens based on a single-share design centred at a wavelength of 1280 nm, utilising an area fraction of $p = 1/11$. The corresponding PSFs were measured at incident wavelengths of 1230 nm, 1280 nm, and 1330 nm, with their longitudinal intensity profiles shown in **Fig. 4c**. As expected, chromatic aberration results in focal spot shifts along the optical axis by approximately 2 mm across the 100 nm spectral range, although diffraction-limited focusing performance is retained at each wavelength. In contrast, the 11-share synthetic achromatic metalens maintains a consistent focal position across the entire spectral band from $\lambda_1 = 1200$ nm to $\lambda_{11} = 1400$ nm (**Fig. 4d**), with a focal shift of only ~1/40 relative to the single-share reference device. The associated transverse PSFs are presented in **Fig. 4e**. To highlight the diffraction-limited focusing performance, the synthetic metalens exhibits an average SR of $0.869 \pm 0.073$ across the spectral window, as exemplified by the PSF profile at 1200 nm, which closely matches theoretical predictions (**Fig. 4f**). To construct the synthetic achromatic metalens, we aligned the T-shaped elements within each nonlocal meta-pixel to have identical in-plane orientations for a given phase value, resulting in a large aperture diameter of 8.1 mm. In contrast, using differently oriented T-shaped elements within a single meta-pixel enables local phase variation, thereby increasing the total number of phase pixels and enhancing focusing efficiency (Supplementary Materials S3D and **Fig. S7**). Importantly, this synthetic strategy introduces no additional design complexity compared to its single-wavelength metalens counterpart. Beyond achromatic focusing, our synthetic metasurface design also enables multifunctional displays, integrating 5 wavelength-specific functions into a single device. This allows simultaneous displays of a holographic image, two distinct OAM modes, and two independent gratings at 5 discrete wavelengths (**Fig. S8**).

We next implemented the disorder-enabled synthetic strategy to realise a polarimetric imaging metasurface capable of single-shot detection of arbitrarily structured light fields with high spatial resolution. The design workflow is illustrated in **Fig. 5a**. Three spatial shares are randomly sampled within the metasurface (area fraction $p = 1/3$) and encoded with three sets of polarisation-selective gratings. Each set projects one of three orthogonal polarisation bases—horizontal/vertical linear polarisation (LPH/LPV), diagonal/anti-diagonal linear polarisation (LPD/LPA), and left-/right-handed circular polarisation (CPL/CPR)—into distinct directions in the momentum ($k$) space. To realise selective responses for these polarisation bases, we used three sets of meta-pixels (**Fig. 5b**), with the full meta-atom library provided in **Fig. S9**. The synthetic metasurface was fabricated using a-Si nanopillars on a quartz substrate. SEM images of the fabricated device are shown in **Fig. 5c**, with the three meta-pixel sets highlighted in false colours (as annotated in **Fig. 5b**).

The polarimetric imaging performance was first validated by characterising 100 targeted polarisation states distributed across the Poincaré sphere. Details of the optical characterisation setup are provided in Supplementary Materials S4 (**Figs. S10 and S11**). The experimentally retrieved polarisation results closely match the targets, yielding a negligible mismatch of 0.039±0.017 (**Fig. S12**). Leveraging our disorder sampling strategy, simulations further reveal that the minimum area required to achieve reliable polarisation-resolved projections can be reduced to as small as 3.2 μm x 3.2 μm (corresponding to 8 x 8 meta-atoms) (**Fig. S13**). This capability unlocks single-shot polarimetric imaging of arbitrarily structured light fields with complex, spatially varying polarisation states—typically requiring high-resolution polarisation detection. To demonstrate this functionality, we used the synthetic metasurface to directly image complex vector beams in a single acquisition. Unlike the Gaussian-shaped *k*-space projections observed for homogeneous polarisations, structured vector beams produce distinct spatial signatures across the three orthogonal polarisation bases. For example, the k-space patterns of radially (**Fig. 5d**) and azimuthally (**Fig. 5e**) polarised beams, along with other complex vector fields (**Fig. S14**), exhibit spatially resolved features that precisely align with polarisation camera results, demonstrating the capability of the metasurface in resolving spatially inhomogeneous vector beams. We further applied the metasurface to characterise optical skyrmions—topological light fields with particle-like spatial textures and non-trivial polarisation structure [48]. **Figure 5f** presents an optical skyrmion formed by superimposing orthogonal linear polarisations with distinct OAM modes ($l_H = 1$, $l_V = 0$). The experimentally measured topological number of 0.987 shows excellent agreement with the simulated value of 0.999. Two additional vector beams were similarly characterised using the same device (**Figs. S15-S17**), matching well with their simulation and polarisation camera results. As such, our synthetic metasurface enables previously unattainable single-shot, high-spatial-resolution polarimetric imaging of arbitrarily structured light fields.

**Conclusion**

In summary, we have demonstrated that disorder can be harnessed to more efficiently utilise metasurface area and substantially enhance functional density. Using a synthetic design strategy, we implement disordered sampling of functionally distinct meta-pixels, each independently addressable via multiple degrees of freedom of light—such as wavelength, polarisation, and OAM. More specifically, our disorder-enabled synthetic achromatic metalens integrates 11 wavelength-specific lens profiles, realised through nonlocal meta-pixels supporting qBICs. These meta-pixels, composed of a minimal number of unit cells required to support sharp resonant behaviour, exhibit an experimentally measured Q factor of ~150. This design ensures that each wavelength operates according to its designated phase profile, while suppressing transmission at off-resonant wavelengths to reduced inter-share crosstalk. Our synthetic achromatic metalens achieves broadband achromatic focusing across 1200 nm to 1400 nm, with only a ~1/40 focal shift as compared to its chromatic counterpart. Notably, our approach requires only the simple design of conventional lenses at individual wavelengths—standing in sharp contrast to state-of-the-art methods, which necessitate sophisticated compensation schemes to simultaneously tailor both phase and dispersion at the level of single meta-atoms [19-25, 45].

Beyond lensing, we incorporate polarisation-selective meta-atoms into a synthetic polarimetric imaging metasurface, enabling the implementation of polarisation-selective diffraction gratings corresponding to three sets of orthogonal polarisations. This synthetic metasurface achieves functional isolation of individual gratings in *k* space. Using this platform, we have experimentally demonstrated single-shot polarimetric imaging of arbitrary polarisation states across the Poincaré sphere with high precision (0.039±0.017), along with high spatial resolution (~3.2 μm) essential for imaging complex vector beams, which is previously unattainable. Altogether, this disorder-enabled synthetic metasurface platform allows for seamless integration of diverse photonic functionalities within a single aperture, representing a substantial advance toward high-density, multifunctional photonic devices.

Although our present demonstrations focus on "one-dimensional" functionalities (either wavelength- or polarisation-addressable), the synthetic approach is readily extendable to high-dimensional control encoded across multiple degrees of freedom. Following the OAM holography principle [18], helical phase profiles can be used to develop the OAM selectivity in a synthetic hologram. Meanwhile, metasurfaces capable of simultaneously manipulating both polarisation and wavelength are readily achievable using our method. Moreover, the synthesis framework supports hierarchical integration: achromatic and polarimetric metasurfaces, for example, can be unified into a single device that functions simultaneously as an achromatic lens and a polarimeter. Crucially, the simplicity of our design process—enabled by functional isolation in either real or *k*-space—underpins this scalability; without it, the design complexity would rapidly become intractable.

**Methods**

*Numerical simulation:*

Simulations for sampled lensing and synthetic achromatic lenses in Figure 2 were performed with angular spectrum methods. Different fractions (*p)* of total pixel number of 1000 by 1000 were utilised for calculations. In Figures 2a-2d, 5a, the pixel sizes are enlarged for the visualisation. In disordered cases, the SRs are averaged from 5 configurations with different random seeds. For Figure 3, the number of 'T'-shaped qBIC meta-pixels and their resonances were analysed using COMSOL Multiphysics. The unit-cell library for Figure 5 was constructed through parametric sweep performed with a custom MATLAB program based on rigorous coupled-wave analysis.

*Fabrication:*

Metasurface sample fabrication employs standard electron-beam lithography (EBL) and deep-reactive ion etching (DRIE) processes. For metalens samples, a 100 nm a-Si layer was deposited on a 500 μm quartz substrate using plasma-enhanced chemical vapor deposition (PECVD, Oxford PlasmaPro100). ZEP520A e-beam resist was then spin-coated at 3000 rpm and baked at 180 °C for 3 minutes on a hotplate. A 20 nm chromium (Cr) layer was sputtered to ensure surface conductivity. Electron exposure was performed using a 100 kV EBL system (Raith, EBPG 5000). The Cr layer was removed with a chromium etchant before development. The sample was immersed in n-amyl acetate for 60 seconds, rinsed with isopropanol and deionized water, and dried with nitrogen gas. The samples were etched in DRIE using $SF_6$ and $C_4F_8$ mixed gases, followed by a 3-minute oxygen plasma treatment to remove residual ZEP resist. For polarimetric metasurface samples, a 500 nm a-Si layer was deposited in PECVD. PMMA resist (A4) was first spin-coated at 3000 rpm followed by soft-baking at 180 °C for 3 minutes. Then, an ultrathin layer of conducting polymer coated (Allresist, Electra 92) to enhance surface conductivity. The same EBL exposure process was followed afterwards. Development was done in a MIBK:IPA (1:3) mixture for 60 seconds. Afterwards, a 30 nm Cr layer was deposited so that it could be used as hard masks after liftoff in acetone. Target pattern was finally transferred to the a-Si layer using DRIE, and the Cr mask was removed via wet etching.

*Metalens and Metasurface sample characterisation:*

Optical characterisation of metalens and polarimetric metasurfaces was performed in a transmission configuration with setup details schematically shown in Supplementary Materials (**Figs. S5 and S10**).

**Acknowledgments:** This work was performed in part at the Melbourne Centre for Nanofabrication (MCN) in the Victorian Node of the Australian National Fabrication Facility (ANFF). The authors thank Ondřej Červinka for the assistance on the polarimetric imaging experiment.

**Funding**: The authors acknowledge the following funding support.

H. R.: Australian Research Council grant (DE220101085, DP220102152);

S. A. M.: Australian Research Council grant DP220102152, Lee Lucas Chair in Physics;

A.F.: Thanks to the Oppenheimer Memorial Trust, the CSIR/NRF Rental Pool Programme, and the South African Quantum Technology Initiative.

**Author contributions:** H. R., C. Liu and C. Li conceived the project. C. Liu modelled the disorder and function density part in Figure 2. C. Li and H. R. modelled achromatic and polarimetric metasurfaces for Figure 3-5. C. Li fabricated all the metasurface samples. C. Li, H. R and H. Y designed and performed the experiments for achromatic metalens and polarimetric metasurfaces in Figure 5d-f. C. P. and A. F. characterised vector fields, Skyrmion samples and plotted related data. C. Li, C. Liu processed the data and plotted the figures. C. Li, C. Liu and H. R. wrote the first draft of the manuscript. A. F., S. A. M., contributed to discussion of the results. H. R. supervised the project.

**Competing interests:** The authors declare that they have no competing interests.

**Data and materials availability:** All data are available in the main text or the Supplementary Materials.

**Supplementary Materials**

Methods

Supplementary Text

Figs. S1 to S17

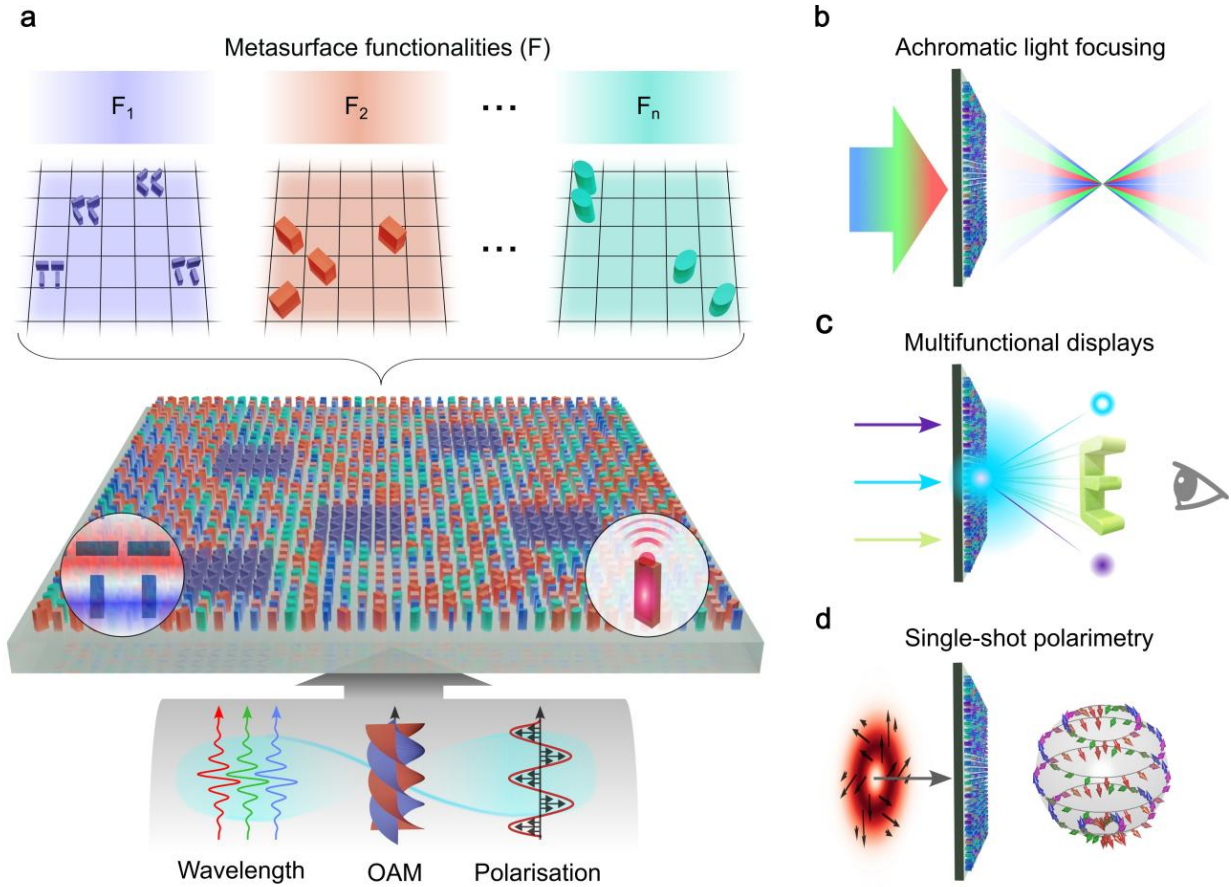

**Figure 1**. **Schematic illustration of a synthetic metasurface platform with substantially enhanced functional density.** (**a**) A synthetic metasurface based on the disordered spatial arrangement of functionally distinct meta-pixels (top)—comprising either nonlocal- (left inset) or local-type (right inset) phase elements—supports multiple optical functionalities. These meta-pixels can be selectively addressed via distinct wavelengths, orbital angular momentum (OAM), and polarisations, or their combinations. Different colours of the meta-pixels indicate distinctly assigned functions. (**b**-**d**) Demonstrated implementations of the synthetic metasurface platform: (b) achromatic light focusing; (c) multifunctional displays at distinct wavelengths; and (d) Single-short polarimetric imaging of arbitrarily structured light fields.

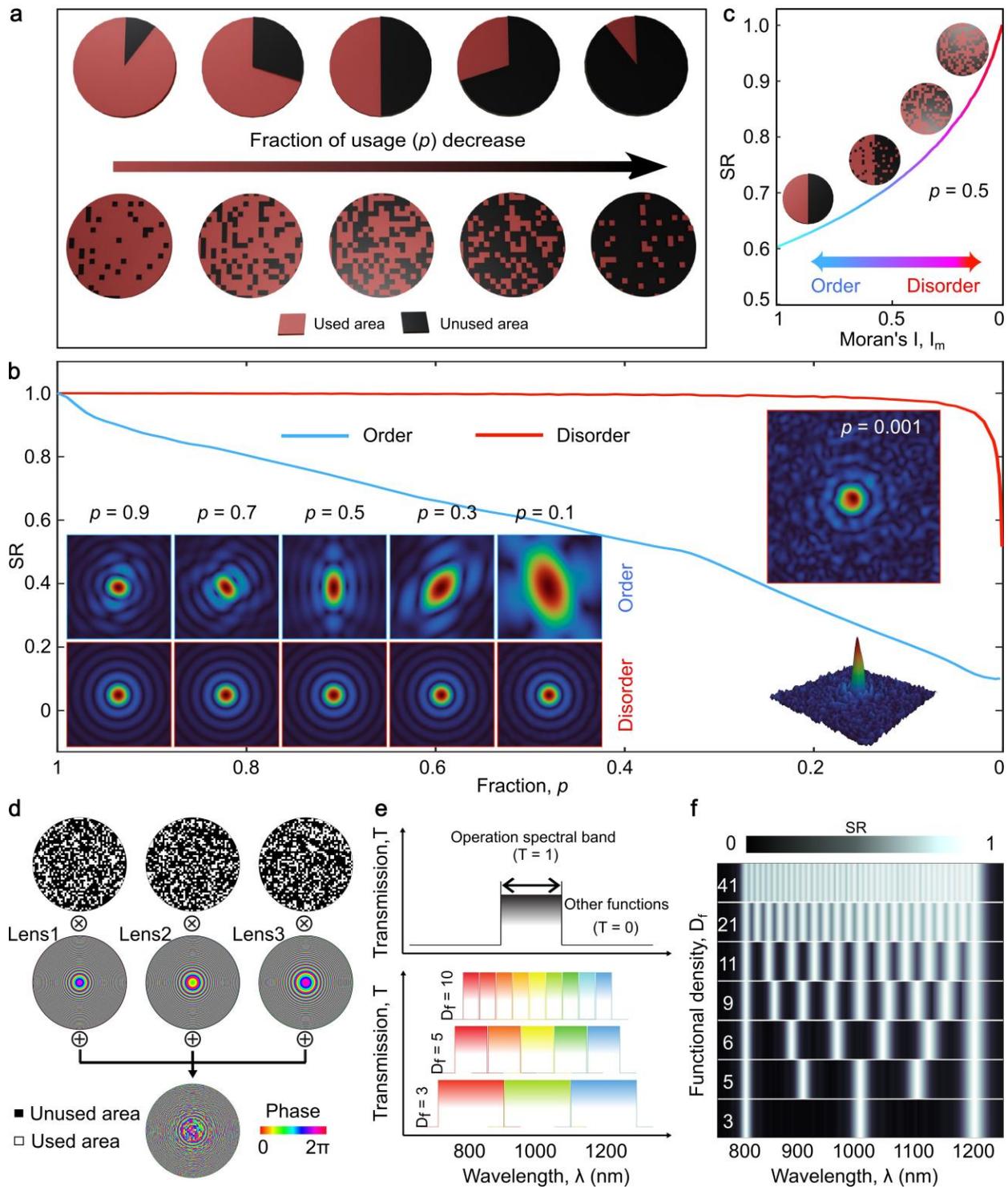

**Figure 2. Disorder-enhanced functional density and its application in synthetic metasurfaces.** (**a**) Comparison of two strategies for area utilisation in single-function metasurfaces: sector-based segmentation (ordered, top) and random sampling (disordered, bottom). As the area usage fraction *p* decreases, progressively less area is allocated to the target function. (**b**) Strehl ratio (SR) of a metalens as a function of *p* for the ordered (blue curve) and disordered (red curve) configurations. The left inset displays the corresponding point spread functions (PSFs) - ordered (upper row) and disordered (lower row). The disordered approach enables minimal area usage without significant degradation of focusing performance. The rightmost inset shows the PSF for $p=0.1\%$, where the focal spot remains discernible. (**c**) Evolution of the SR during the order-to-disorder transition, quantified by the Moran's spatial autocorrelation index, where higher values indicate greater spatial ordering. Insets illustration of the transition from ordered to disordered area utilisation. (**d**) Synthesis of a multifunctional metasurface from individual single-function metasurfaces. A functional density of $D_f = 3$ is used. Each individual function corresponds to a hyperbolic phase profile for light focusing at a specific wavelength, implemented using a disordered sampling strategy with non-overlapping pixel regions among different function shares. The bottom row illustrates the resulting synthetic phase profile, constructed by superimposing the individual phase profiles corresponding to each wavelength channel. (**e**) Schematics of an ideal bandpass filter for a single function (top), and a synthetic achromatic metalens comprising multiple crosstalk-free bandpass filters (bottom). (**f**) Achromatic focusing performance of the synthetic metalens, characterised by Strehl ratio (SR) values across a broadband spectrum, evaluated as functions of functional density and wavelength.

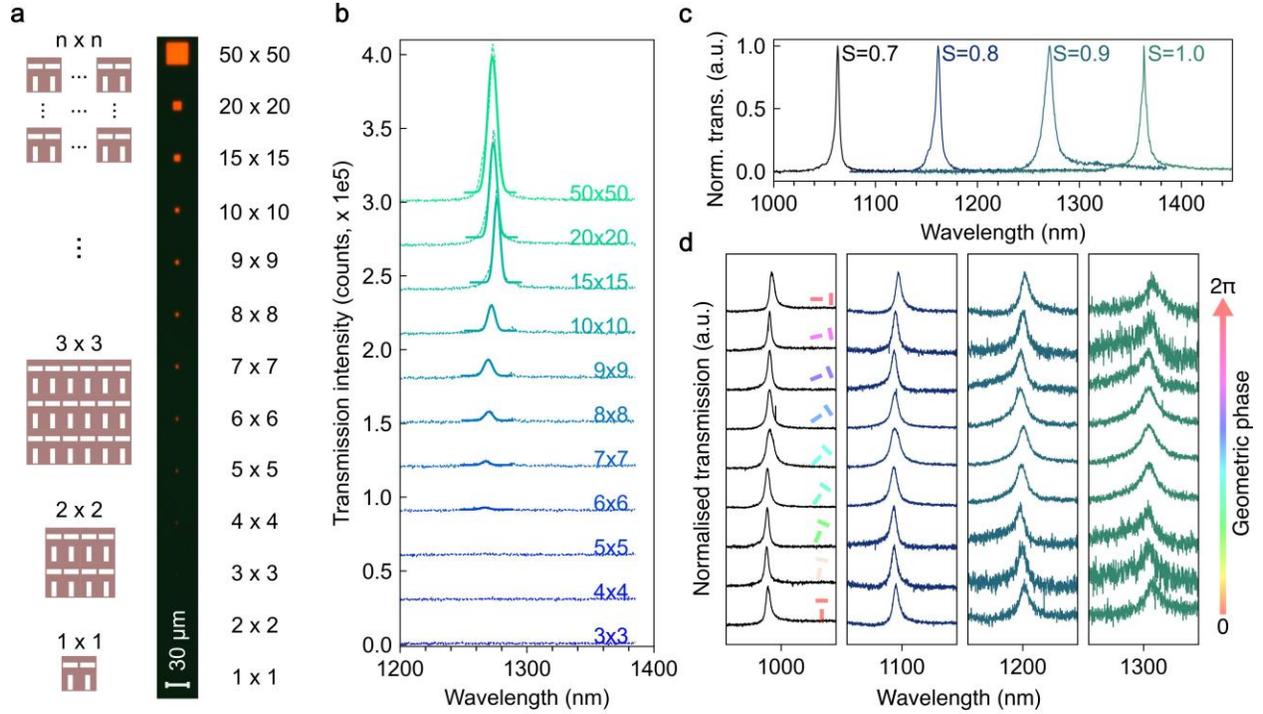

**Figure 3. Resonance characterisation of T-shaped meta-pixels supporting quasi-bound states in the continuum (qBICs), showing the dependence on unit cell number and robustness against in-plane rotational variations.** (**a**) Schematics of meta-pixels composed of T-shaped unit-cell arrays, alongside an optical image of the fabricated sample under cross-circular polarisation illumination and collection. Each unit cell contains four rectangular elements configured into two "T" shapes within a square pixel geometry. Scale bar: 30 μm. (**b**) Stacked transmission spectra corresponding to the meta-pixels shown in (a), with fitted resonance peaks overlaid. (**c**) Resonance tuning enabled by varying the geometric scaling factor (*S*). (**d**) Robust resonance behaviour observed for meta-pixels with different in-plane rotational angles spanning a complete 0–2π geometric phase range. All meta-pixels maintain a fixed area of approximately 8.1 μm × 8.1 μm.

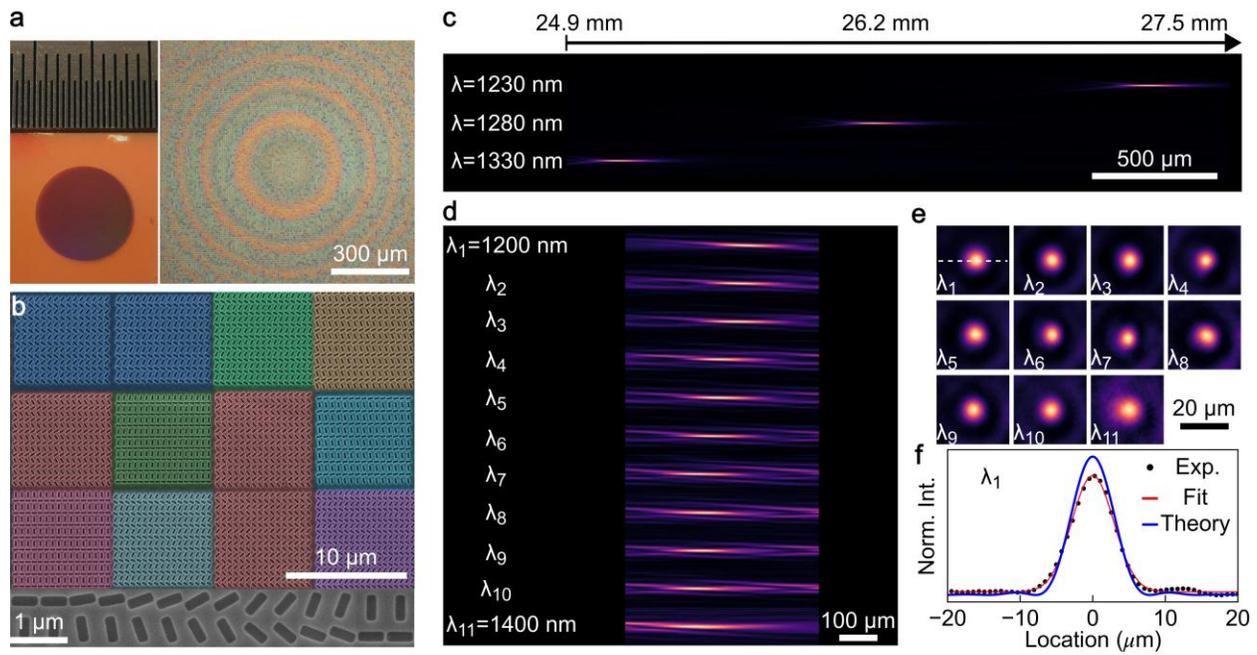

**Figure 4. Fabrication and optical characterisation of a synthetic achromatic metalens.** (**a**) Optical images of the fabricated synthetic metalens with an 8 mm diameter on a quartz substrate. (**b**) SEM image of zoom-in meta-pixels, with functionally distinct pixels highlighted in false colours. The rotated structures used for phase encoding are magnified for clarity in the bottom row. (**c**) Measured longitudinal point spread function (PSF) of a single-share chromatic metalens (without achromatic design), characterised at 1230 nm, 1280 nm, and 1330 nm. (**d**) Stacked longitudinal PSFs of the synthetic achromatic metalens across the spectral range from 1200 nm to 1400 nm, exhibiting a focal shift of only 1/40 compared to its chromatic counterpart. (**e**) Transverse PSF maps of the synthetic achromatic metalens at 11 distinct wavelengths, demonstrating diffraction-limited focusing across the spectral range. (**f**) Line profile of the PSF in (e) at 1200 nm, overlaid with the theoretical diffraction-limited prediction. Both peak areas are normalised to 1 to show the diffraction-limited focusing (Strehl ratio ~ 0.87)

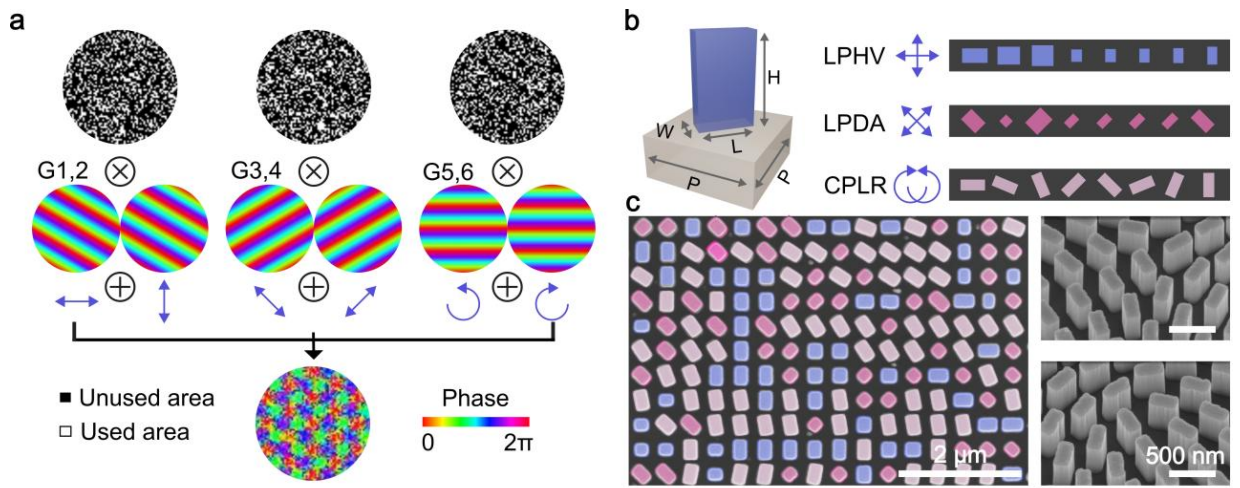

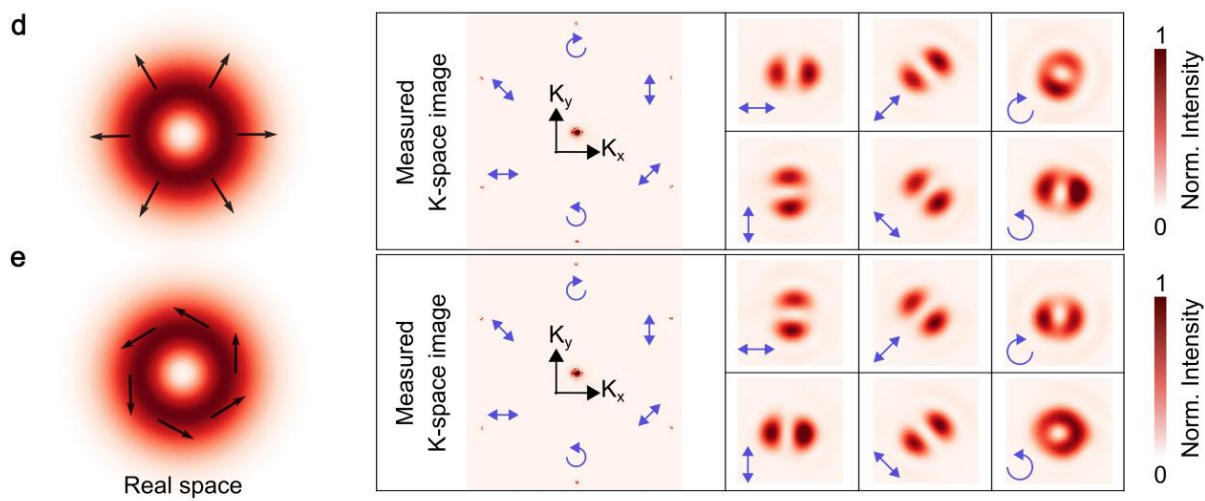

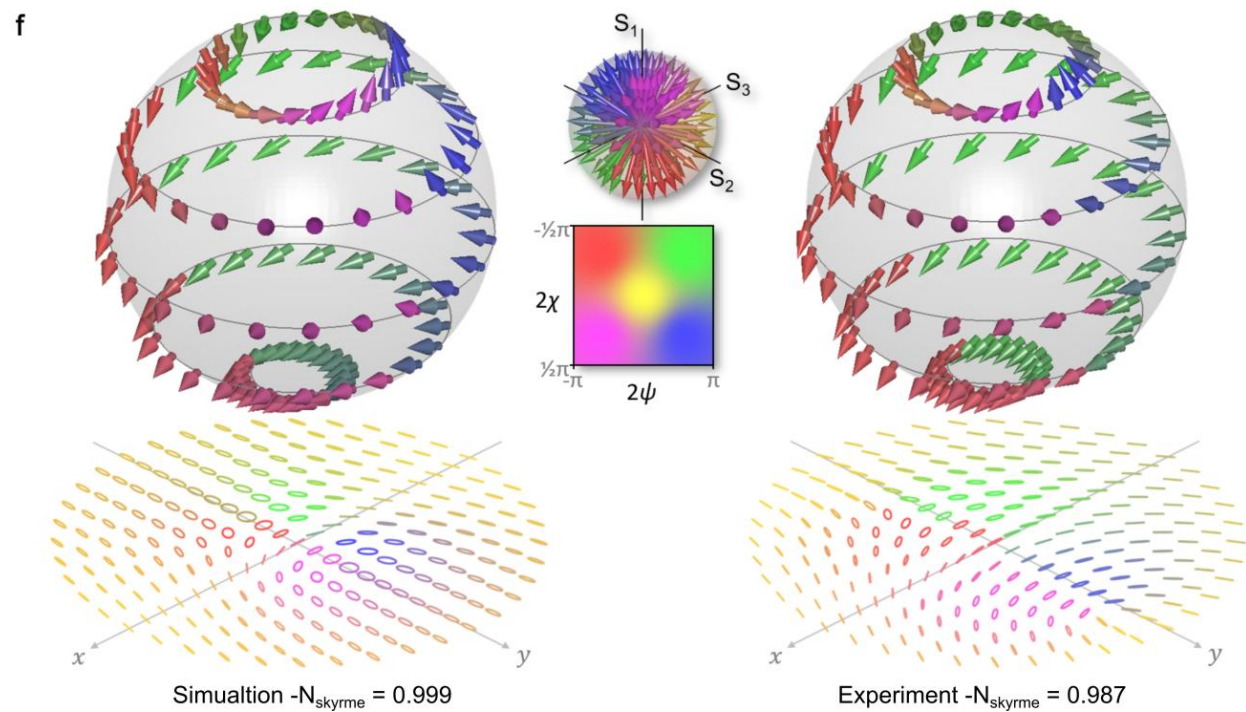

Simualtion -$N_{skyrme}$ = 0.999

Experiment -$N_{skyrme}$ = 0.987

**Figure 5. Synthetic polarimetric metasurfaces for single-shot polarimetry of both spatially homogeneous and arbitrarily structured vector fields.** (**a**) Synthesis of the polarimetric metasurface aperture from three distinct disordered pixel shares, each implementing a polarisation-selective diffraction grating. Each share deflects incident light with orthogonal polarisation bases into opposite directions. (**b**) Schematic of three sets of meta-pixels based on amorphous silicon nanopillars on a quartz substrate, with representative dimensions annotated. Examples include designs for linearly and circularly polarised fields: horizontally (vertically) oriented nanopillars provide phase control for linear polarisations (LPHV), while ±45° rotations encode phase responses for diagonal and anti-diagonal polarisations (LPDA). A uniform subwavelength half-wave plate, varied by in-plane rotation, imparts geometric phase shifts to shape left- and right-circular polarisations into opposite directions (CPLR). (**c**) SEM image of the fabricated synthetic metasurface, with functionally distinct pixel shares shown in false colours. (**d** and **e**) Single-shot polarimetric imaging of complex vector beams with radial (d) and azimuthal (e) polarisation distributions. Left panels show real-space beam profiles with polarisation vectors. Middle and right panels display characteristic *k*-space intensity distributions captured by the metasurface, revealing six diffraction orders corresponding to six polarisation bases. (**f**) Simulated (left) and experimental (right) stereographic projections of a polarisation Skyrmion with a Skyrme number of 1, exhibiting rich vector textures. Experimental data acquired in a single shot using the synthetic polarimetric imaging metasurface.